\documentclass[conference]{IEEEtran}
\IEEEoverridecommandlockouts
\usepackage[numbers,comma,sort&compress]{natbib}
\usepackage{flushend}
\usepackage{amsmath,amssymb,amsfonts}
\usepackage{algorithmic}
\usepackage{graphicx}
\usepackage{hyperref}
\usepackage{textcomp}
\usepackage[table,xcdraw]{xcolor}
\usepackage{colortbl}
\definecolor{gold}{rgb}{1.0, 0.84, 0.0}
\definecolor{silver}{rgb}{0.75, 0.75, 0.75}
\definecolor{green1}{rgb}{0.51, 0.68, 0.59}
\usepackage{soul}
\usepackage{booktabs} 
\usepackage{multirow}
\usepackage[linesnumbered,ruled,lined,commentsnumbered]{algorithm2e}
\usepackage[inkscapelatex=false]{svg}
\def\BibTeX{{\rm B\kern-.05em{\sc i\kern-.025em b}\kern-.08em
    T\kern-.1667em\lower.7ex\hbox{E}\kern-.125emX}}
    \graphicspath{{./figures/}}
\begin{document}

\title{RTSR: A Real-Time Super-Resolution Model for AV1 Compressed Content\\

\thanks{The authors appreciate the funding from Netflix Inc., University of Bristol, and the UKRI MyWorld Strength in Places Programme (SIPF00006/1).}
}

\author{
Yuxuan Jiang$^1$,
Jakub Nawała$^1$,
Chen Feng$^1$,
Fan Zhang$^1$,
Xiaoqing Zhu$^2$,
Joel Sole$^2$,
and David Bull$^1$ \\
$^1$ \textit{Visual Information Laboratory, University of Bristol, Bristol, BS1 5DD, UK}\\
$^1$ \textit{\{yuxuan.jiang, jakub.nawala, chen.feng, fan.zhang, dave.bull\}@bristol.ac.uk}\\
$^2$ \textit{Netflix Inc., Los Gatos, CA, USA, 95032}\\
$^2$ \textit{\{xzhu, jsole\}@netflix.com}\\
}

\maketitle

\begin{abstract}
Super-resolution (SR) is a key technique for improving the visual quality of video content by increasing its spatial resolution while reconstructing fine details. SR has been employed in many applications including video streaming, where compressed low-resolution content is typically transmitted to end users and then reconstructed with a higher resolution and enhanced quality. To support real-time playback, it is important to implement fast SR models while preserving reconstruction quality; however most existing solutions, in particular those based on complex deep neural networks, fail to do so. To address this issue, this paper proposes a low-complexity SR method, RTSR, designed to enhance the visual quality of compressed video content, focusing on resolution up-scaling from a) 360p to 1080p and from b) 540p to 4K. The proposed approach utilizes a CNN-based network architecture, which was optimized for AV1 (SVT)-encoded content at various quantization levels based on a dual-teacher knowledge distillation method. submitted to the AIM 2024 Video Super-Resolution Challenge, specifically targeting the Efficient/Mobile Real-Time Video Super-Resolution competition. It achieved the best trade-off between complexity and coding performance (measured in PSNR, SSIM and VMAF) among all six submissions. The code will be available soon.
% The code is available at \url{https://github.com/YuxuanJJ/ECCVAIM2024}.
\end{abstract}


\section{Introduction}
\label{sec:intro}

Visual content has become increasingly prominent in today's digital landscape \cite{bull2021intelligent}. The widespread use of digital devices has led to a significant increase in the consumption of video content across various applications such as live streaming, broadcasting, video conferencing, and surveillance. These video-centric applications now represent a substantial portion, approximately 80\%, of global internet traffic \cite{r:cisco2}.

\begin{figure}[!t]
    \centering

    \includegraphics[width=1\linewidth]{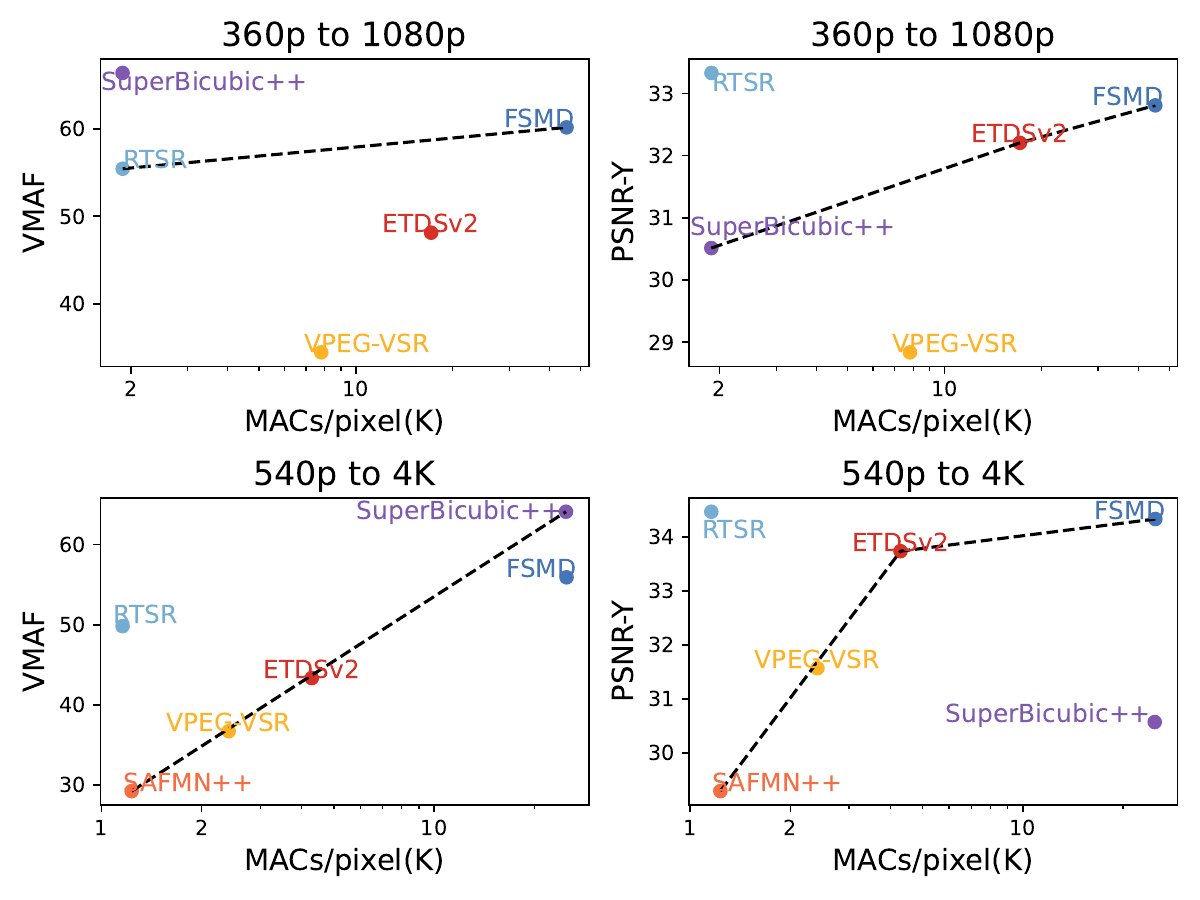}

    % \begin{minipage}[b]{0.485\linewidth}
    %     \centering
    %     \centerline{\includegraphics[trim={5cm 2cm 1.2cm 2cm},clip, width=.98\linewidth]{x3.pdf}}
    %     \text{$\times$3}\vspace{.1cm}
    %     \end{minipage}
    % \begin{minipage}[b]{0.485\linewidth}
    %     \centering
    %     \centerline{\includegraphics[trim={5cm 2cm 1.2cm 1cm},clip, width=.98\linewidth]{x4.pdf}}
    %     \text{$\times$4}\vspace{.1cm}
    %     \end{minipage}
    
    \caption{Scatter plot for six methods (\ref{sec:com-model}) in terms of the balance between VMAF and PSNR-Y performance and MACs/pixel. The upper left indicates a better performance-complexity trade-off for the corresponding method.}
    \label{fig:radar-chart}
    \vspace{-0.5cm}
\end{figure}

Video compression is one of the most widely researched topics in image and video processing. It plays a crucial role in the trade-off between the (substantial) bit rate required for transmitting immersive, high-quality video content and the limited network bandwidth available. Over the past few decades, the efficiency of video codecs has seen notable advances, with the development of video coding standards including  H.266/VVC \cite{VVCSoftware_VTM} and AOM AV1 \cite{AV1}. These standards are, however, all based on the same rate-distortion optimization framework as their predecessors (e.g., HEVC \cite{h265HEVC} and VP9 \cite{VP9}). As such they may not be able to fulfill the significantly growing demand associated with emerging media formats, in particular when higher spatial resolutions are adopted.

Recently, inspired by advances in deep learning, numerous learning-based video coding methods have emerged \cite{dvc, kwan2023hinerv, mimt, canfvc, kwan2024nvrc, li2024neural}. Among these solutions, an important class focuses on the use of deep neural networks to enhance individual coding tools within conventional coding architectures \cite{yan2018convolutional, zhang2020enhancing, ma2020cvegan, ma2020mfrnet, ViSTRA3, li2024video, huo2024towards, ibraheem2024enhancing}. Although these approaches have shown significant promise, delivering evident coding gains, they are also associated with high complexities, with slow run-times and large memory requirements. Moreover, for many of these methods, simple distortion-based loss functions, e.g., mean squared error (MSE) and L1, are utilized in the training process, despite their relatively poor correlation with perceived visual quality. As a result, they often fail to produce results with optimal perceptual quality \cite{ma2020cvegan}.

In this context, this paper describes a low-complexity super-resolution approach, RTSR, specifically optimized for AV1 compressed content. Inspired by the EDSR block \cite{lim2017enhanced}, this method is designed to up-scale 360p compressed videos to 1080p resolution and 540p to 4K in real-time. To further enhance super-resolution performance, a knowledge distillation strategy, proposed in \cite{jiang2023compressing, jiang2024mtkd} has been applied; a perceptually-inspired Generative Adversarial Network (GAN) architecture, CVEGAN \cite{ma2020cvegan} and EDSR\_baseline \cite{lim2017enhanced} were used as dual-teacher models. Moreover, a perceptually-inspired loss function, developed in \cite{ma2020cvegan}), is employed in the training and optimization processes in order to produce results with improved perceptual video quality. This approach has been integrated within the SVT-AV1 \textit{version 1.8.0} video codec and was submitted to the Grand Challenge in Real Time Video Super-Resolution (ECCV 2024 AIM (Advances in Image Manipulation) Workshop (Team BVI-RTVSR)) \cite{conde2024aim}. As illustrated in Figure \ref{fig:radar-chart}, RTSR offers the best trade-off between complexity and performance - it exceeds the Pareto front of the other five solutions in three of four cases with different quality metrics (PSNR and VMAF) and complexity combinations.

% In summary, the contributions of this paper are concluded as follows.

% \begin{itemize}
%     \item We proposed a real-time SR model for enhancing AV1 compressed video content .
%     \item We employ a dual-teacher knowledge distillation training strategy for improving the performance of the low-complexity model.
%     \item The experimental results show that the proposed model achieves the best balance between performance and complexity. 
% \end{itemize}

The rest of the paper is organized as follows. Section \ref{sec:PA} describes the proposed RTSR method, the integrated coding framework, and the training process. The experiment setup is detailed in Section \ref{sec:E}. The coding results are then analyzed in Section \ref{sec:RD}. Finally, Section \ref{sec:c} concludes the paper and outlines future works.

% \section{Related Work}
% \label{sec:PA}

\section{Proposed Algorithm}
\label{sec:PA}

\begin{figure}[t]
    \centering
    \includegraphics[width=0.9\linewidth]{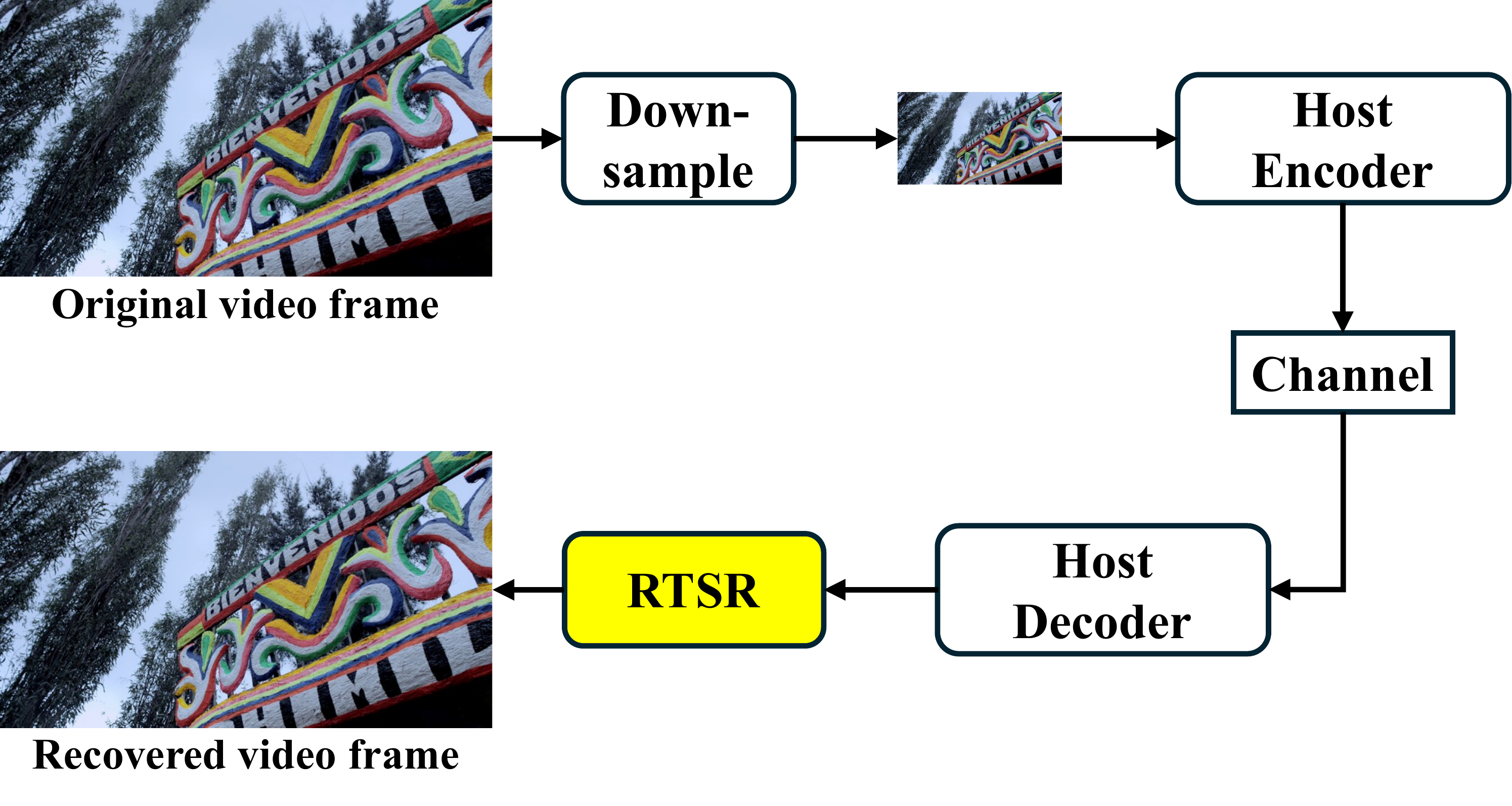}
    \caption{\small The proposed coding framework, with an RTSR module.}
    \label{fig:coding_frame}
    \vspace{-0.5cm}
\end{figure}

The proposed coding framework is illustrated in Figure \ref{fig:coding_frame}. Prior to encoding, the original YUV input 1080p video is downsampled by a factor of 3 or 4 using a Lanczos filter. SVT-AV1 v1.8.0 \cite{AV1} serves as the Host Encoder used to compress low-resolution videos. At the decoder, the low-resolution video stream is decoded and the proposed RTSR approach is applied to reconstruct the full resolution video content while enhancing its perceptual quality. Details of the network architecture and the training process are described below.

\begin{figure}[t]
    \centering
    \includegraphics[width=1\linewidth]{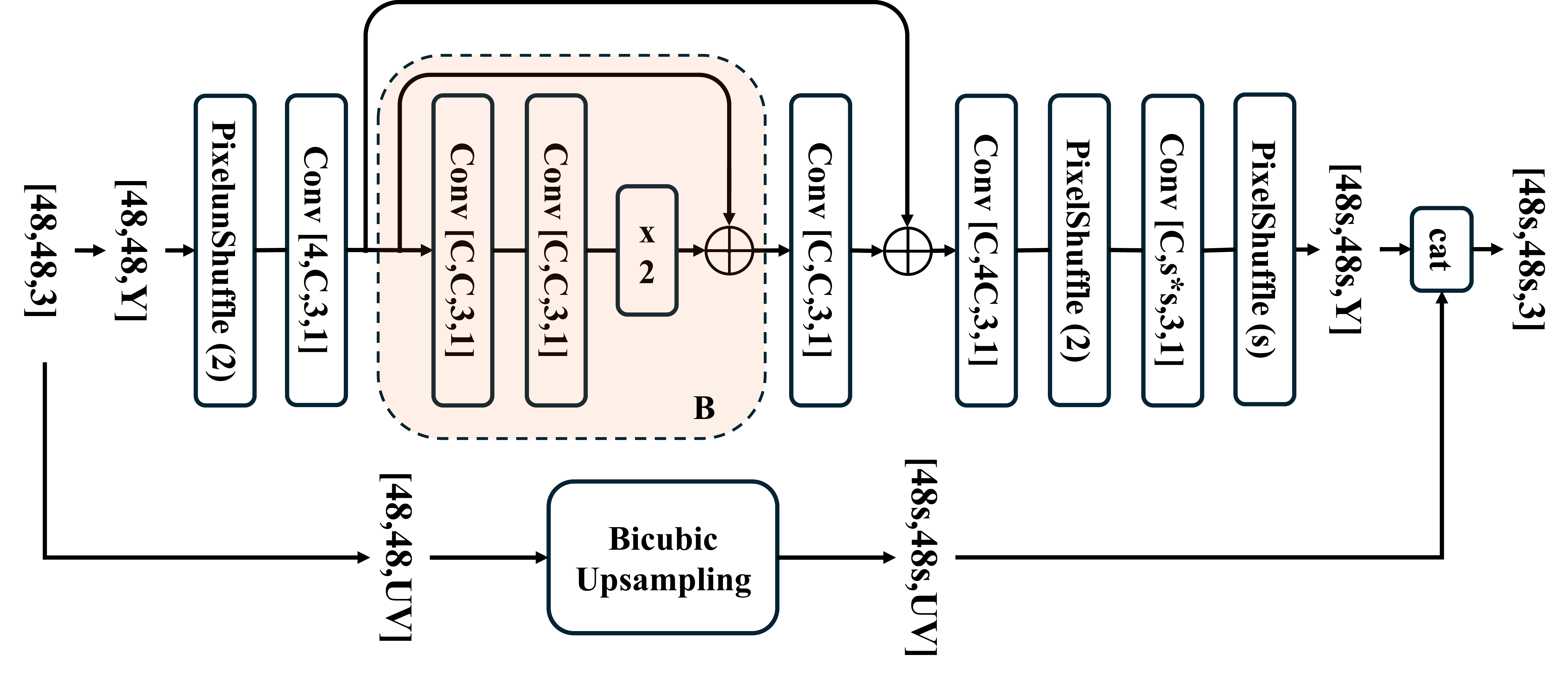}
    \caption{\small The network structure of the proposed RTSR model. Factor $s$ represents the upsampling factor. Here block number B is set to 3, and channel number C is 24.}
    \label{fig:network}
    \vspace{-0.5cm}
\end{figure}

\subsection{Network Architecture}

The network architecture, shown in Figure \ref{fig:network}, takes a 48$\times$48 YCbCr 4:2:0 compressed image block as input, which is first up-sampled using a nearest neighbor (NN) filter to 48$\times$48 YCbCr 4:4:4 \cite{ma2020cvegan}.  The output is a 144$\times$144 image block (for the $\times$3 up-scaling) in YCbCr 4:4:4 format, which is then converted into YCbCr 4:2:0 format, targeting its original uncompressed full resolution version. For the $\times$4 up-scaling, a factor of 4 is used in the final PixelShuffle layer, producing a 192$\times$192 image block as output. 

Knowledge distillation (KD) has emerged as a promising technique for improving the performance of low-complexity super-resolution networks \cite{jiang2024mtkd, he2020fakd}. We use this approach in the proposed RTSR model which learns from two teacher models: a perceptually-inspired network for compressed video enhancement, CVEGAN\cite{ma2020cvegan}, and the EDSR\_baseline \cite{lim2017enhanced}.

In order to meet real-time requirements, a PixelunShuffle layer is also used before convolutional layers to greatly reduce the number of operations. The main body of the network consists of B identical blocks, each with a ReLU layer and two consecutive convolutional layers, which are highlighted in Figure \ref{fig:network}. The up-sampling module (after B convolutional blocks) consists of a two-step PixelShuffle operation. UV channels are up-sampled by a bicubic filter and concatenated with the Y-channel recovered by the CNN model.

\subsection{Training Configuration}

The training of the proposed model is performed in two stages, described in Algorithm \ref{alg:training}. In the first stage, the student model is trained from scratch, with a combined perceptual loss function, proposed in \cite{ma2020cvegan}.
\begin{equation}
    \mathcal{L}_{p} = 0.3 \mathcal{L}_\mathit{L1} + 0.2 \mathcal{L}_\mathit{SSIM} + 0.1 \mathcal{L}_\mathit{L2} + 0.4 \mathcal{L}_\mathit{MS-SSIM}
\end{equation}
In the same stage, two teacher models, CVEGAN and EDSR\_baseline, are also trained from scratch with identical training content. We followed the training configurations described in the source publications  \cite{ma2020cvegan, lim2017enhanced}.

In the second stage, the pre-trained RTSR model is used as the student model, which learns from its pre-trained teachers, CVEGAN and EDSR\_baseline using the following loss function:
\begin{equation}
    \mathcal{L}_\mathit{total} = \alpha\mathcal{L}_\mathit{Lap}(I_\mathit{stu}, I_\mathit{gt}) + \Sigma \mathcal{L}_\mathit{Lap}(I_\mathit{stu}, I_\mathit{tchr}),
    \label{Ltotal}
\end{equation}
where $\mathcal{L}_{Lap}(I_{stu}, I_{gt})$ denotes the original loss between the ground truth $I_{gt}$ and the student model’s prediction $I_{stu}$, and $\alpha$ is a tunable weight.  $\mathcal{L}_{Lap}(I_{stu}, I_{tchr})$ represents the loss between the student $I_{stu}$ and the teacher’s predictions $I_{tchr}$. Here $\mathcal{L}_{Lap}$ is the Laplacian loss \cite{niklaus2018context}. We have tested several commonly used loss functions, including $\mathit{L1}$, $\mathit{L2}$, $\mathit{MSE}$, $\mathit{SSIM}$, $\mathit{MS-SSIM}$, and found that Laplacian loss performs the best in terms of various quality metrics including PSNR and VMAF.

The RTSR implementation is based on PyTorch version 1.10 \cite{paszke2019pytorch}. We used the following configurations during training:
Adam optimization \cite{kingma2014adam} with the hyper-parameters: $\beta_{1} = 0.9$ and $\beta_{2} = 0.999$; batch size of 16; 200 training epochs (100 for both stage 1 and 2); initial learning rate of $10^{-4}$; weight decay of 0.1 for every 50 epochs. The hyperparameter $\alpha$ in eq. (\ref{Ltotal}) is set to 0.1, following \cite{morris2023st}.  The block number B is 3, and channel number C is 24, which results in a complexity below 2K MACs/pixel.

\begin{algorithm}[t]
\small
    \SetKwInOut{Input}{Input}
    \SetKwInOut{Output}{Output}
    \SetKwInOut{Output}{Output}
	\SetAlgoLined

        \Input{Training datasets $\mathcal{D}$ contain cropped original sequences $y$ and compressed sequences pairs $x$}
        \Input{Quantisation parameters $\mathcal{Q}=\{31,39,47,55,63\}$}
        \Input{Scaling factors $\mathcal{F}=\{\times3,\times4\}$}
        \Output{Trained student model $S_{\times F}$.}

	\BlankLine
	\For{$F \in \mathcal{F}$}{
	       \textbf{Stage1:} \\
                Train CVEGAN $T_1$, EDSR\_baseline $T_2$, RTSR $S_{\times F}$ separately with loss function $\mathcal{L}_{p}$ from scratch.
            
                \textbf{Stage2:} \\
                Transfer knowledge from $T_1$ and $T_2$ to $S_{\times F}$. \\
                \ForEach{batch of data $(x, y)$ in $D$}{
                    Compute teacher1 predictions: $z_{T_1} = T(x)$\;
                    Compute teacher2 predictions: $z_{T_2} = T(x)$\;
                    Compute student predictions: $z_S = S_{\times F}(x)$\;
                    Compute distillation loss: $\Sigma \mathcal{L}_\mathit{Lap}(I_\mathit{z_S}, I_\mathit{z_{T_i}})$\;
                    Compute standard loss: $\mathcal{L}_\mathit{Lap}(I_\mathit{z_S}, I_\mathit{y})$\;
                    Compute total loss: $\mathcal{L}_\mathit{total}$\;
                    Update student model $S_{\times F}$ using gradient descent\;
                }
            % Well-trained student model $S_{\times F}$.
        }
	\caption{The training strategy used for optimization RTSR.} 
 \label{alg:training}
\end{algorithm}

\section{Experiment Setup}
\label{sec:E}
\subsection{Training Content}

The RTSR, CVEGAN, and EDSR\_baseline models were optimized using the training database described in Sec. \ref{sec:Exp-data}. In addition to the LDV3 videos provided as part of the AIM Grand Challenge \cite{conde2024aim}, original sequences from the BVI-DVC database \cite{ma2021bvi} and the BVI-AOM database \cite{nawala2024bvi} were also employed. BVI-DVC contains 800 diverse uncompressed video sequences at various resolutions, and has been employed as a training database for MPEG JVET to optimize neural network-based coding tools for VVC, while BVI-AOM includes more content with complex structures (e.g., fire, water, or plasma) and artistic intent (e.g., action movie like face close-ups).

\begin{figure}[t]
        \scriptsize
    \centering
    \begin{minipage}[b]{0.495\linewidth}
        \centering
        \centerline{\includegraphics[width=.98\linewidth]{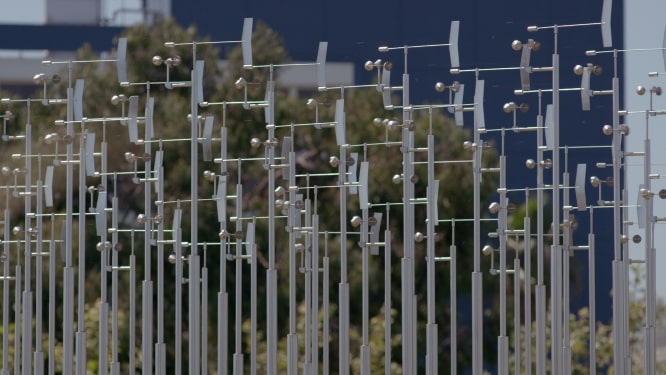}}
        \text{Chimera-WindAndNature}\vspace{.1cm}
        \end{minipage}
    \begin{minipage}[b]{0.495\linewidth}
        \centering
        \centerline{\includegraphics[width=.98\linewidth]{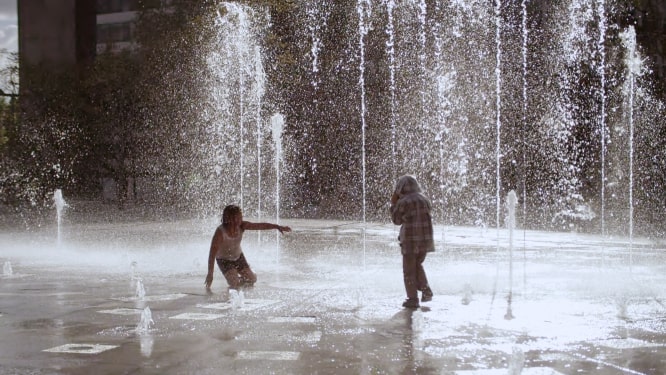}}
        \text{NETFLIXElFuente}\vspace{.1cm}
            \end{minipage}

    \begin{minipage}[b]{0.495\linewidth}
        \centering
        \centerline{\includegraphics[width=.98\linewidth]{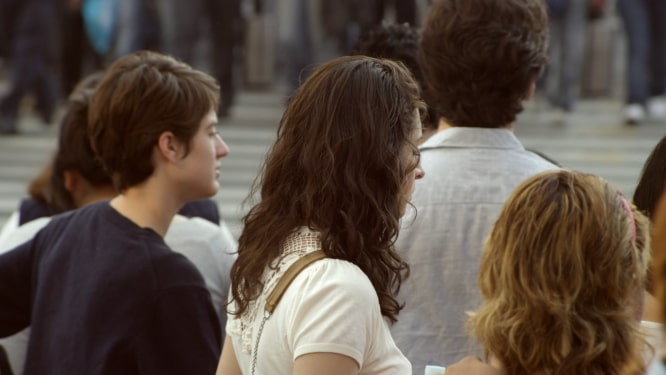}}
        \text{NETFLIXElFuente90s}\vspace{.1cm}
            \end{minipage}
    \centering
    % \begin{minipage}[b]{0.183\linewidth}
    %     \centering
    %     \centerline{\includegraphics[width=.98\linewidth]{Sintel-Dragons.png}}
    %     \text{Sintel-Dragons}\vspace{.1cm}
    %         \end{minipage}
    \begin{minipage}[b]{0.495\linewidth}
        \centering
        \centerline{\includegraphics[width=.98\linewidth]{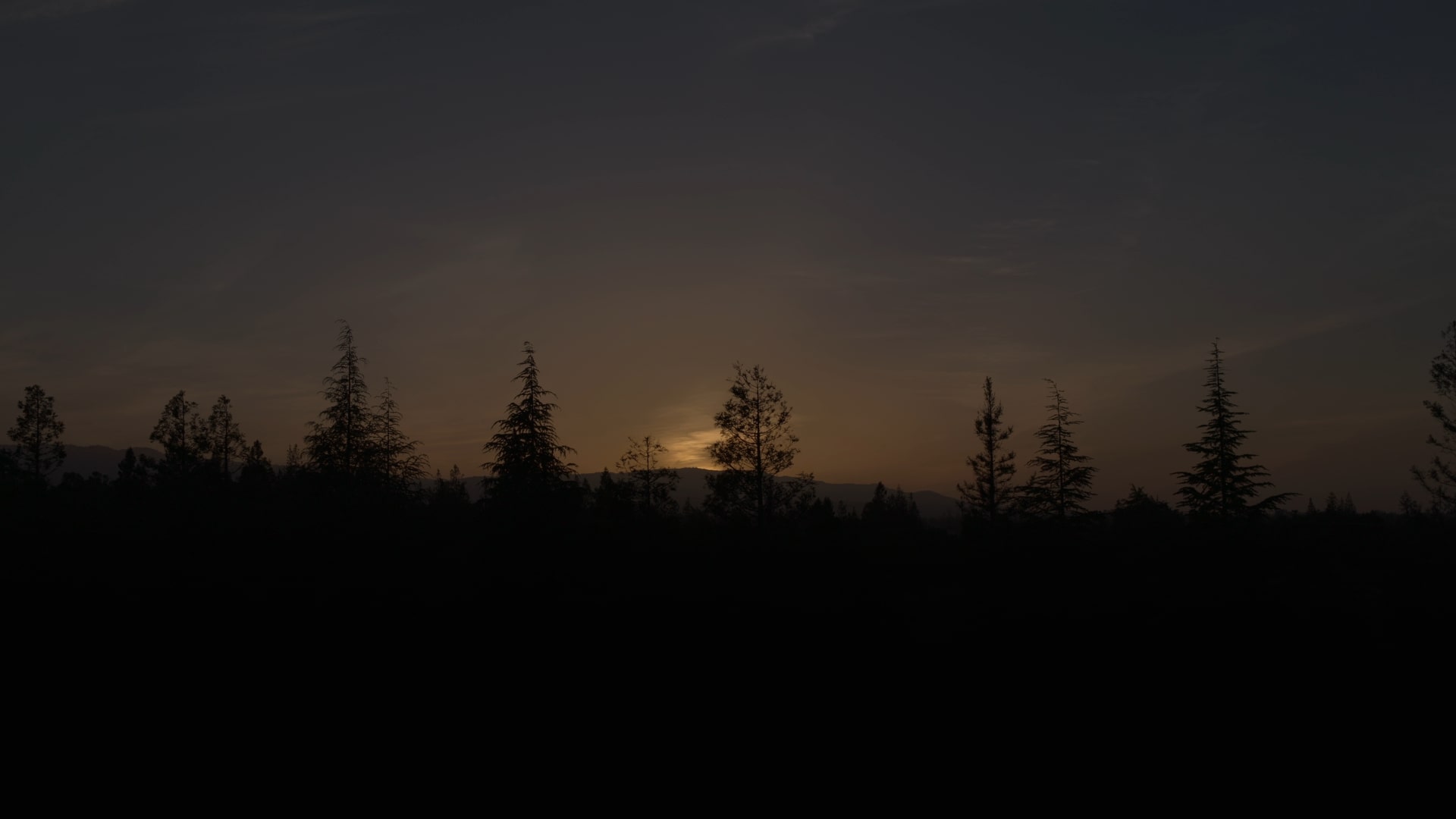}}
        \text{Sparks-Sunrise}\vspace{.1cm}
    \end{minipage}
    
    \caption{Frame samples from the high-quality test set videos. The original videos are in 4K resolution YCbCr 4:2:0 format.}
    \label{fig:testsetexample}
    \vspace{-0.5cm}
\end{figure}

All the original sequences collected were encoded using SVT-AV1 \textit{version 1.8.0}, with five quantization parameter (QP) values (31, 39, 47, 55, and 63). Subsequently, both the compressed sequences and their original counterparts were cropped into 48$\times$48 and 144$\times$144 patches (192$\times$192 for the $\times$4 up-scaling), respectively, and randomly selected for training. Data augmentation techniques such as rotation and flipping were applied to increase content diversity. This results in a total of 92,800 pairs of patches. Based on all the generated training material, we trained a single CNN model for compressed video content with various QPs (for each $\times$3 or $\times$4 up-scaling).

\subsection{Test Datasets and Metrics}
\label{sec:Exp-data}

Nineteen downsampled (using the Lanczos5 filter) sequences, provided by the AIM challenge organizers, have been used to evaluate the effectiveness of the proposed coding framework.  Each sequence contains up to 1799 frames and is compressed with five different QPs, ranging from 31 to 63, after down-sampling. The decoded sequences were also provided by the organizer in mp4 format. Several sample frames are shown in Figure \ref{fig:testsetexample}.

For evaluation, the compressed low-resolution sequences were first converted to the YCbCr 4:4:4 format. The RTSR model processes 16 consecutive low-resolution frames each time. Quality metrics PSNR, SSIM, and VMAF \cite{VMAF} are employed to measure video quality. In addition, model complexity figures, including \#Params(M), MACs(G) and Runtime (ms), are also measured using VideoAI \cite{VideoAI} and reported here. Further information about the AIM 2024 Grand Challenge can be found in \cite{conde2024aim}.

\subsection{Compared Models}
\label{sec:com-model}

The proposed method has been benchmarked against eight reference methods, including EDSR\_baseline \cite{lim2017enhanced}, CVEGAN \cite{ma2020cvegan}, Lanczos5 filter, SuperBicubic++, FSMD, ETDSv2, VPEG-VSR, SAFMN++ \cite{conde2024aim}. The results of the latter six were provided by the AIM 2024 Grand Challenge organizers.

% \begin{table*}[]
%     \centering
%     \begin{tabular}{c|c|c|c|c|c}
%         Input & Track & Train Time (hrs) & \# Params. (M) & MACs (G) & GPU  \\
%         \hline
%          (640,360,3) & $\times$3 & 100 & 0.062 & 3.913 & RTX3090  \\
%          (960,540,3) & $\times$4 & 100 & 0.063 & 9.595 & RTX3090  \\
%     \end{tabular}
%     \caption{Model complexity calculation by using \cite{VideoAI}}.
    
%     \label{tab:my_label}
% \end{table*}

\section{Results and Discussion}
\label{sec:RD}

% \subsection{Test Results}

\begin{table*}[t]
\centering
\caption{Efficient VSR Challenge Benchmark.}
\begin{tabular}{l|c|c|c|c|c|c}
\hline \hline
\textbf{Method} & \textbf{PSNR-Y $\uparrow$} & \textbf{SSIM-Y $\uparrow$} & \textbf{VMAF $\uparrow$} & \textbf{\# Params (M) $\downarrow$} & \textbf{MACs/pixel (K) $\downarrow$} & \textbf{Runtime (ms) $\downarrow$} \\ \hline
\multicolumn{7}{c}{\textbf{Track 1: Mobile, 360p to 1080p ($\times$3)}} \\ \hline
Anchor(Lanczos5)         & 33.123            & 0.9364            & 51.241          & -                 & -         & -         \\ \hline
SuperBicubic++  & 30.513        & 0.9250            & 66.389          & 0.05              & 1.40     & 0.46 (A100)          \\ \hline
FSMD       & 32.808            & 0.9384            & 60.166          & 1.624             & 45.18     & 13.14 (4090)         \\ \hline
ETDSv2     & 32.205            & 0.9333            & 48.127          & 0.136             & 17.15     & 8.6 (A100)         \\ \hline
VPEG-VSR   & 28.836            & 0.8635            & 34.442          & 0.070             & 7.81     & 3.84 (3090)         \\ \hline 
\cellcolor{green1}\textbf{RTSR (ours)}  & \cellcolor{green1}33.329            & \cellcolor{green1}0.9371            & \cellcolor{green1}55.438          & \cellcolor{green1}0.062             & \cellcolor{green1}1.89     & \cellcolor{green1}0.81 (3090)         \\ \hline \hline
\multicolumn{7}{c}{\textbf{Track 2: General, 540p to 4K ($\times$4)}} \\ \hline
Anchor(Lanczos5)         & 34.651            & 0.9577            & 46.049          & -                 & -      & -             \\ \hline
SuperBicubic++  & 30.572        & 0.9416            & 64.112          & 0.398             & 24.92     & 10.77 (A100)        \\ \hline
FSMD       & 34.329            & 0.9591            & 55.920          & 1.599             & 25.02    & 32.33 (4090)           \\ \hline
ETDSv2     & 33.734            & 0.9564            & 43.339          & 0.136             & 4.29   & 8.6 (A100)           \\ \hline
VPEG-VSR   & 31.568            & 0.9111            & 36.704          & 0.077             & 2.41    & 8.56 (3090)           \\ \hline
SAFMN++    & 29.294            & 0.8774            & 29.225          & 0.040             & 1.23   & 8.2 (3090)            \\ \hline 
\cellcolor{green1}\textbf{RTSR (ours)}  & \cellcolor{green1}34.464            & \cellcolor{green1}0.9567            & \cellcolor{green1}49.829          & \cellcolor{green1}0.063             & \cellcolor{green1}1.16     & \cellcolor{green1}2.11 (3090)         \\ \hline \hline
\end{tabular}
\label{tab:challengeresult}
\vspace{-0.2cm}
\end{table*}

% \begin{table}[t]
% \caption{Results  with average PSNR-Y and VMAF across all QPs.}
% \centering
% \begin{tabular}{c| c c c}
% \toprule
%                              & Method & PSNR-Y (dB) & VMAF (score)  \\ \midrule
% \multirow{4}{*}{Track1 ($\times$3)} & EDSR\_baseline & 33.73  & 57.41 \\
%                              & CVEGAN & 33.69 & 57.92 \\
%                              & Lanczos filter & 33.14  & 51.28 \\
%                              & \textbf{Ours}   & \textbf{33.33}  & \textbf{55.44} \\ \midrule
% \multirow{4}{*}{Track2 ($\times$4)} & EDSR\_baseline & 35.32  & 52.16 \\
%                              & CVEGAN & 35.30 & 53.03 \\
%                              & Lanczos filter & 34.66  & 45.92 \\
%                              & \textbf{Ours}   & \textbf{34.46}  & \textbf{49.83}\\
%                              \bottomrule
% \end{tabular}
% \label{tab:performance}
% % \vspace{-0.5cm}
% \end{table}

\begin{figure*}[t]
        \scriptsize
    \centering
    \begin{minipage}[b]{0.3\linewidth}
        \centering
        \centerline{\includegraphics[clip, viewport=120 120 600 450, width=.98\linewidth]{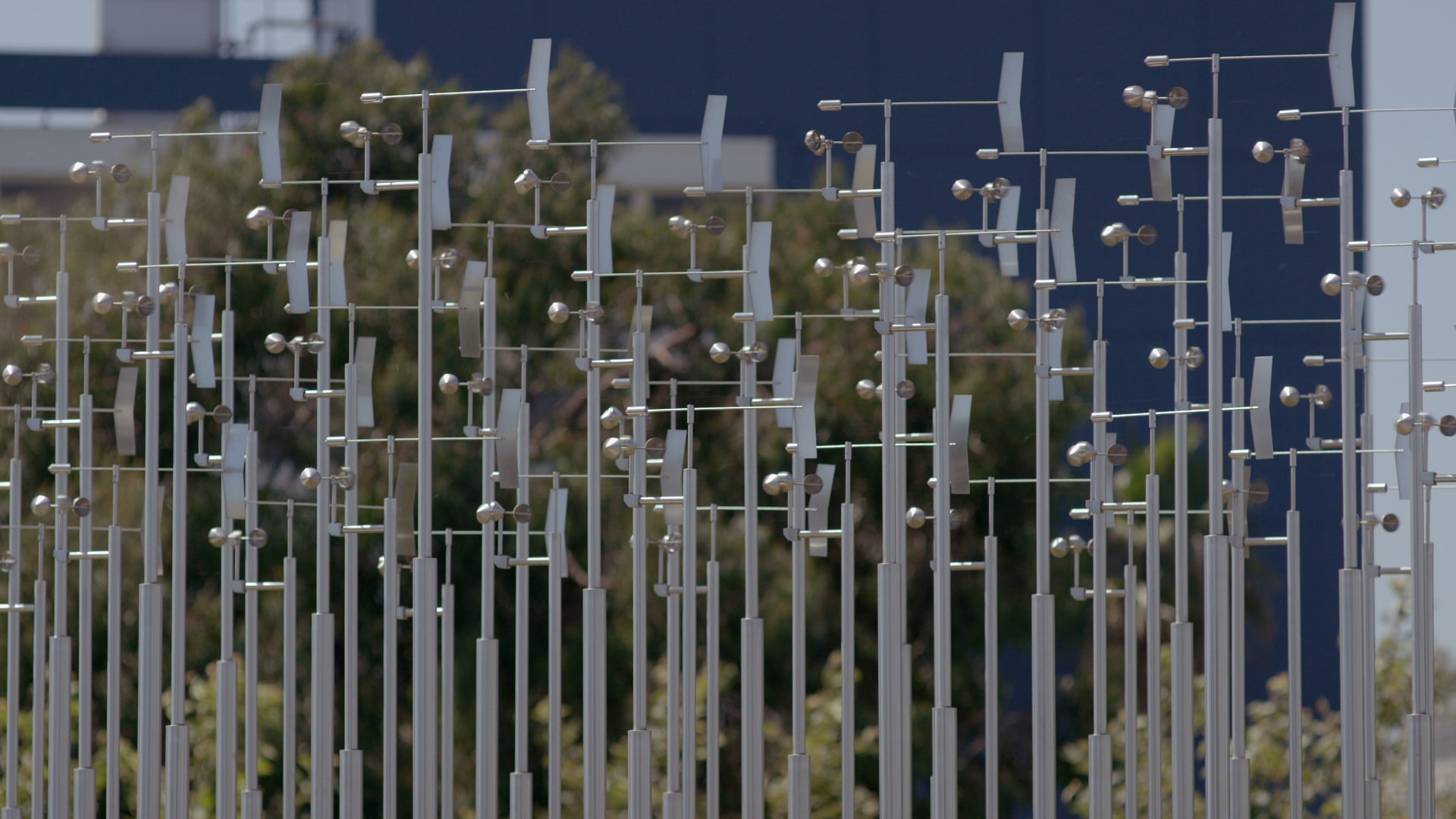}}
        \text{HD HR}\vspace{.1cm}
        \end{minipage}
    \begin{minipage}[b]{0.3\linewidth}
        \centering
        \centerline{\includegraphics[clip, viewport=120 120 600 450, width=.98\linewidth]{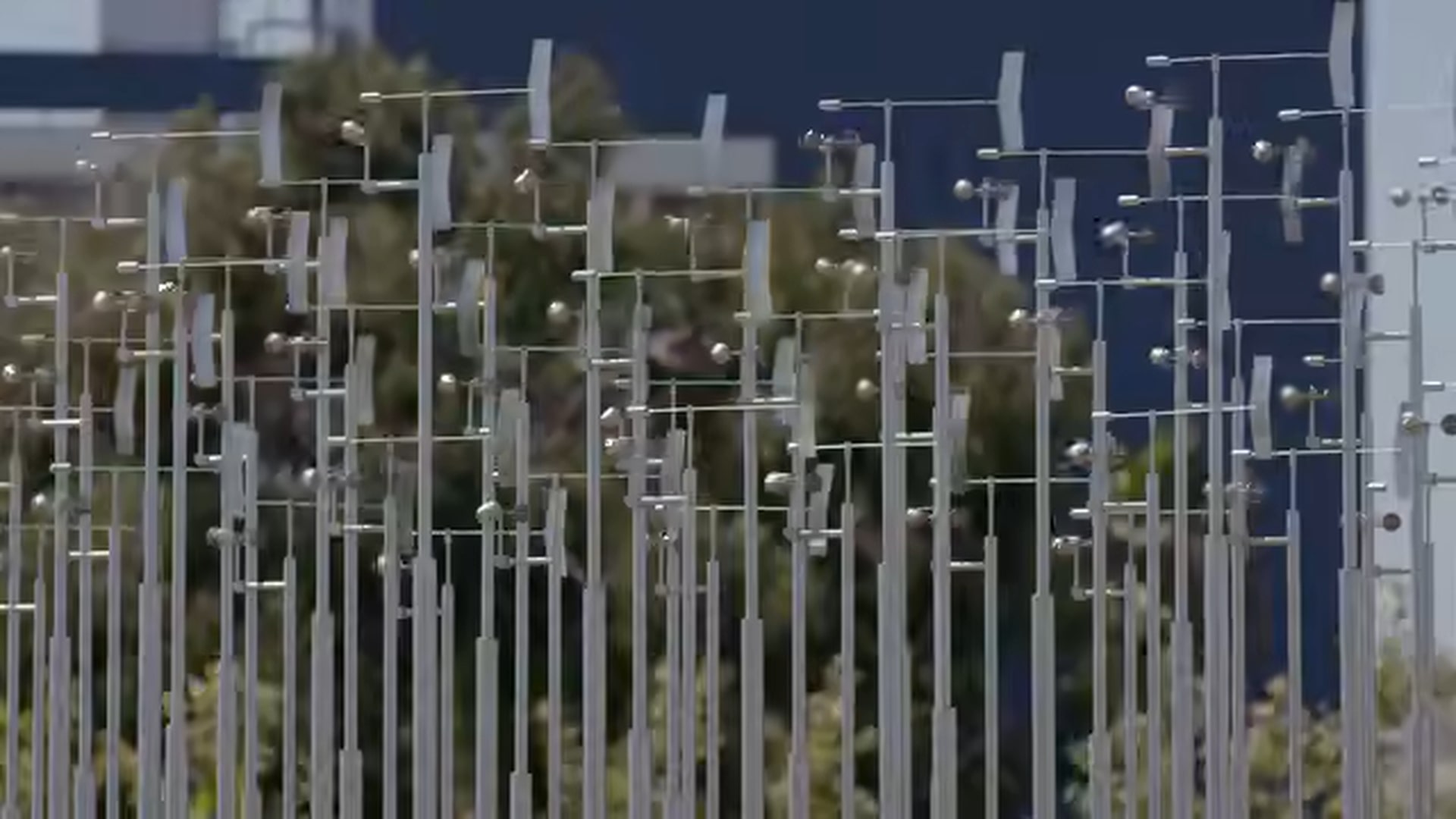}}
        \text{HD Lanczos}\vspace{.1cm}
        \end{minipage}
    \begin{minipage}[b]{0.3\linewidth}
        \centering
        \centerline{\includegraphics[clip, viewport=120 120 600 450, width=.98\linewidth]{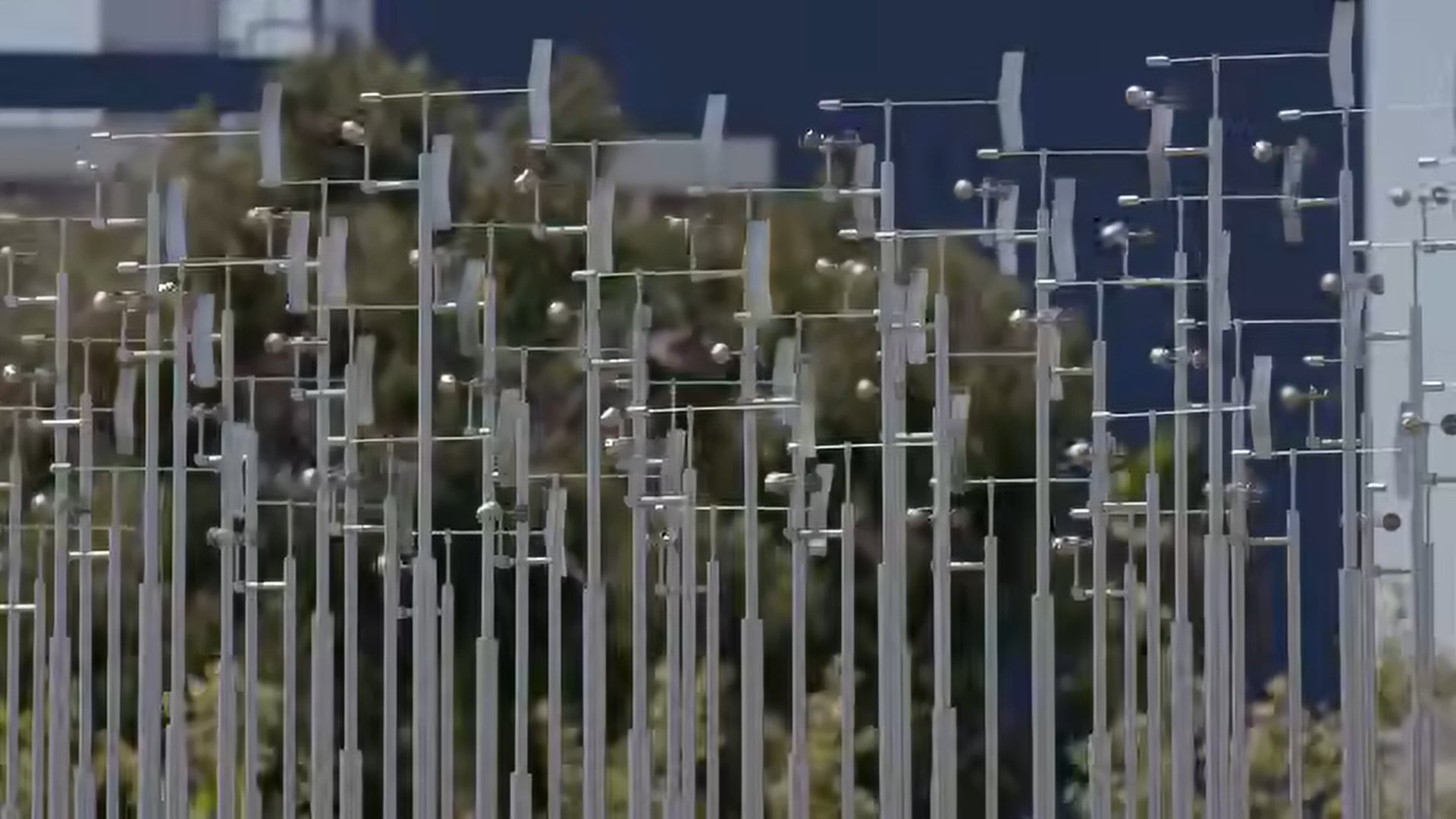}}
        \text{HD SR}\vspace{.1cm}
        \end{minipage}

    \begin{minipage}[b]{0.3\linewidth}
        \centering
        \centerline{\includegraphics[clip, viewport=120 150 600 480, width=.98\linewidth]{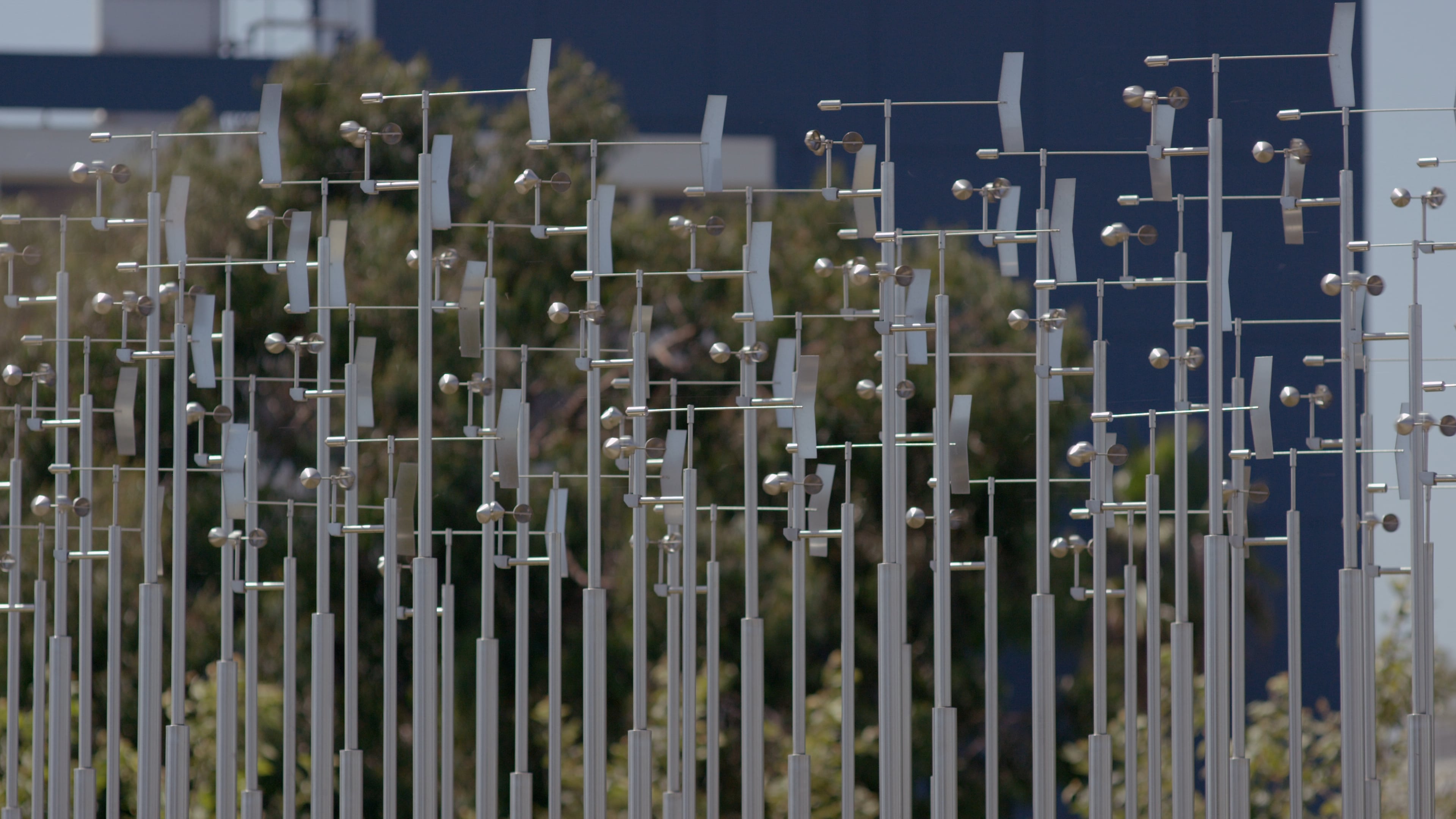}}
        \text{4K HR}\vspace{.1cm}
            \end{minipage}  
    \begin{minipage}[b]{0.3\linewidth}
        \centering
        \centerline{\includegraphics[clip, viewport=120 150 600 480, width=.98\linewidth]{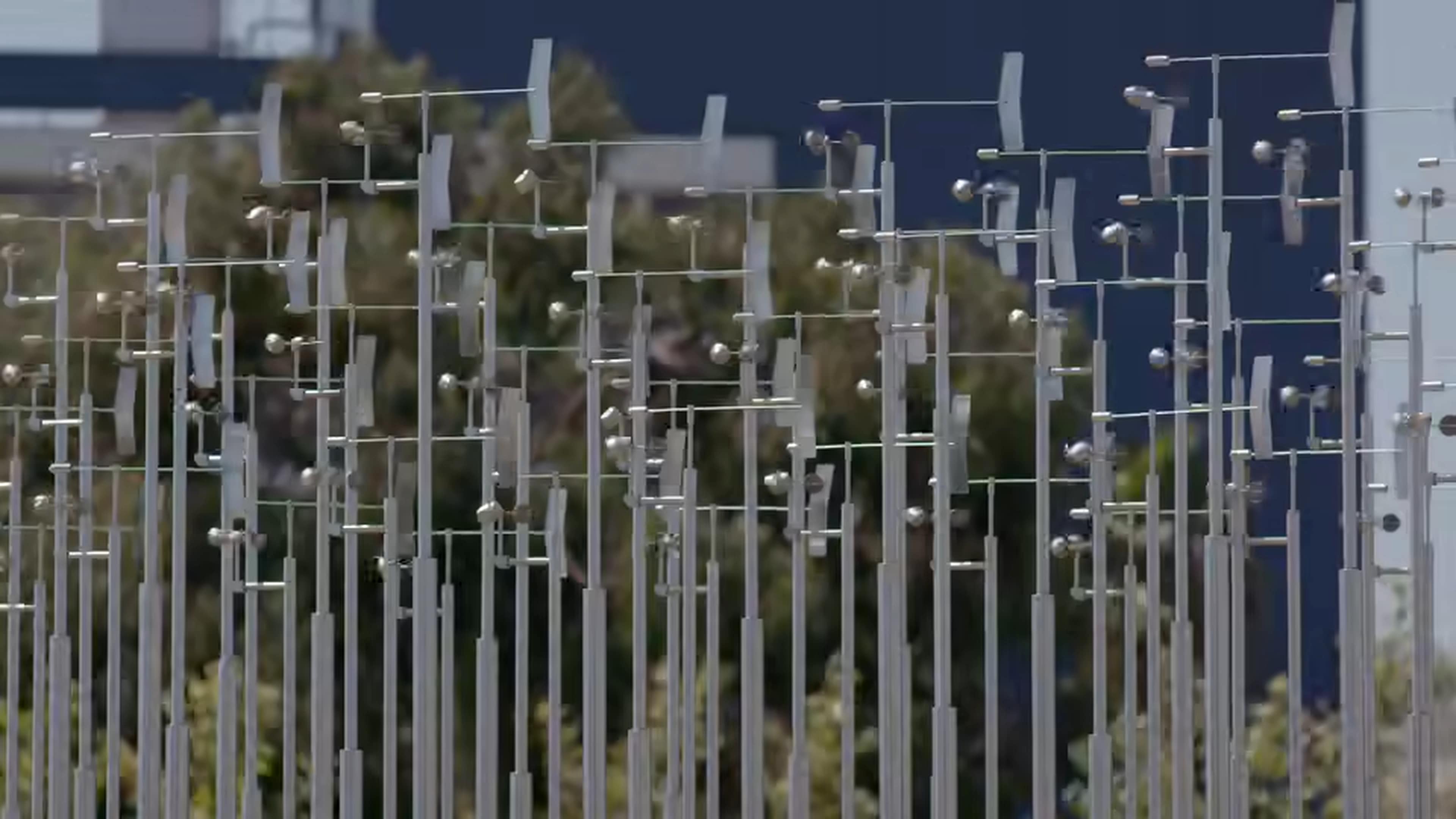}}
        \text{4K Lanczos}\vspace{.1cm}
        \end{minipage}
    \begin{minipage}[b]{0.3\linewidth}
        \centering
        \centerline{\includegraphics[clip, viewport=120 150 600 480, width=.98\linewidth]{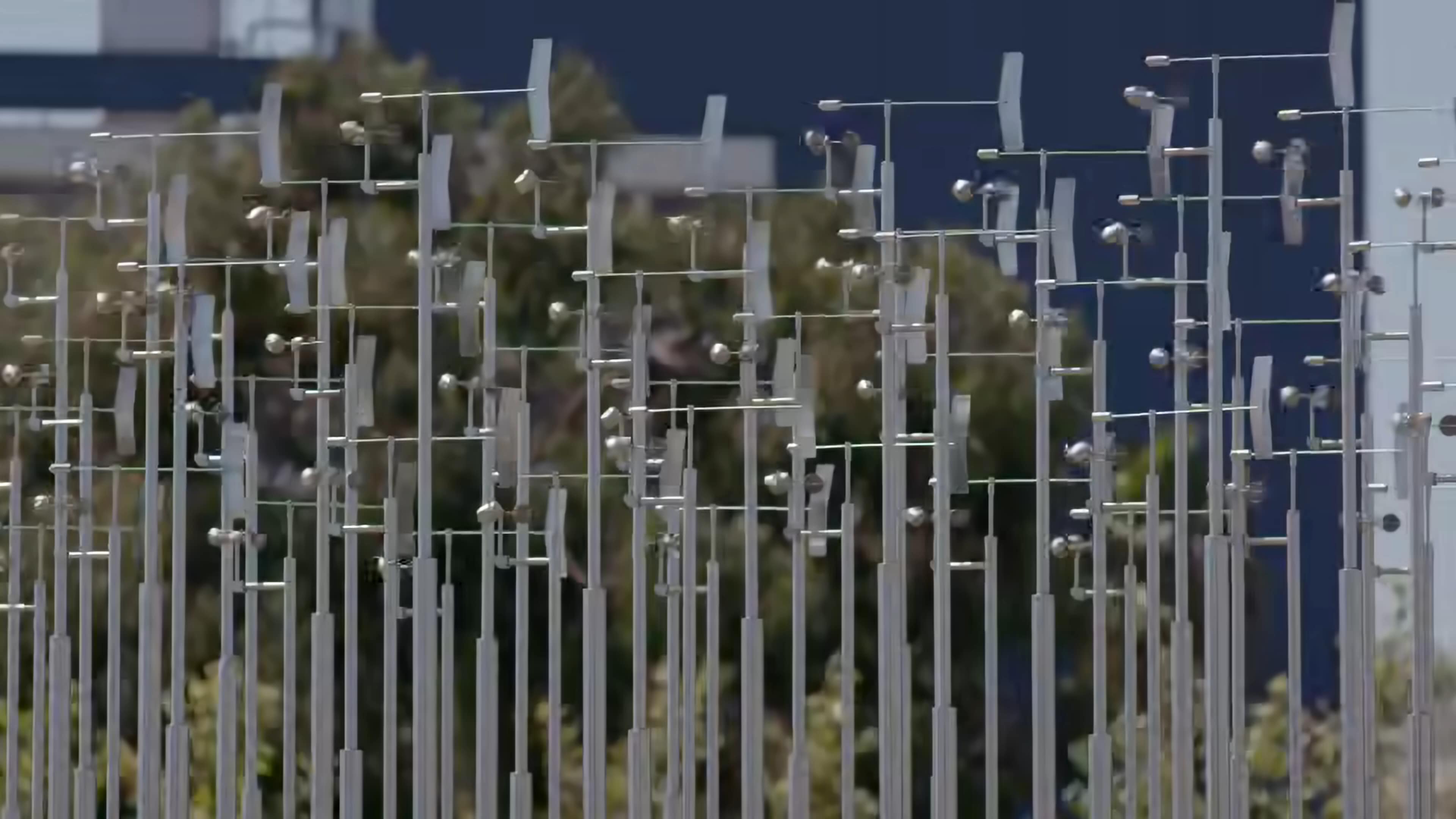}}
        \text{4K SR}\vspace{.1cm}
     \end{minipage}
    \vspace{-0.3cm}
    \caption{Visual comparison: The Chimera-WindAndNature sequence is downsampled and compressed using CRF=55. The ground truth, Lanczos filter upsampling, and RTSR recovery results are shown for comparison.}
    \label{fig:visual}
    \vspace{-0.5cm}
\end{figure*}

Table \ref{tab:challengeresult} summarizes the results of the proposed RTSR method alongside the other five models submitted to the AIM 2024 Grand Challenge \cite{conde2024aim} in terms of the average PSNR-Y, SSIM-Y and VMAF values for all test sequences and rate points (QPs). It can be observed that RTSR offers the best PSNR-Y performance among all six models, with up to a 0.5dB gain over the second best (in Track 1). Although its VMAF performance is relatively lower when directly compared to SuperBicubic++ and FSMD, RTSR provides the best overall trade-off between complexity and performance, as shown in Figure \ref{fig:radar-chart}, where the point corresponding to RTSR is beyond the Pareto front of the other five models in three out of four cases. It is noted that while these metrics have been commonly used to measure compression performance, they are not designed or optimized for super-resolved (in particular AI-based super-resolved) content. Their reliability for this type of content has previously been reported to be lower than that for compressed content \cite{zhou2024database,mackin2018study,mackin2018srqm}. It can also be observed that the processing speed of RTSR is 0.8 ms per frame for the $\times 3$ up-scaling and 2 ms for $\times 4$, which demonstrates the model's ability to handle high-resolution up-scaling tasks with impressive efficiency. 

% \subsection{Visual Comparison}

Figure \ref{fig:visual} presents a visual comparison between RTSR and the Lanczos5 filter (the anchor in the challenge)\footnote{We are unable to obtain results generated by other submitted models.}. It can be seen that RTSR effectively preserves fine details with enhanced visual quality. It not only reduces visible artifacts but also offers a more natural, visually pleasing output, which closely aligns with the original high-resolution sequence. This demonstrates the superior perceptual performance of RTSR compared to traditional upsampling methods.

\section{Conclusion}
\label{sec:c}

In this paper, a low-complexity super-resolution method, RTSR, is presented, which has been designed to upscale 360p SVT-AV1 compressed videos to 1080p and 540p to 4K. The CNN-based network used was specifically optimized for AV1 compressed content using a dual-teacher knowledge distillation training strategy, which enhanced the overall reconstruction performance. The results show its excellent super-resolution performance across diverse content alongside a good trade-off between complexity and performance. Future work should focus on further improving the performance of the model and its application to other coding standards.

\newpage
\small
\bibliographystyle{ieeetr}
\bibliography{refs}

\end{document}